\documentclass[aps,prb,twocolumn,floats,showpacs,superscriptaddress]{revtex4-1}
\usepackage{graphicx,epsfig}
\usepackage{times,bbm}
\usepackage{graphics,dcolumn,bm,float}
\usepackage{amssymb,amsmath,rotate,color}
\usepackage[title,titletoc,toc]{appendix}
\usepackage{mathtools}
\usepackage{booktabs}
\usepackage{tcolorbox}
\usepackage[mathscr]{eucal}
\usepackage{graphicx,epsfig}
\usepackage[pagebackref=false,colorlinks,linkcolor=magenta,citecolor=cyan,urlcolor=magenta]{hyperref}

\begin{document}
\unitlength 1 cm
\newcommand{\be}{\begin{equation}}
\newcommand{\ee}{\end{equation}}
\newcommand{\bea}{\begin{eqnarray}}
\newcommand{\eea}{\end{eqnarray}}
\newcommand{\nn}{\nonumber}
\newcommand{\vk}{\vec k}
\newcommand{\vp}{\vec p}
\newcommand{\vq}{\vec q}
\newcommand{\vkp}{\vec {k'}}
\newcommand{\vpp}{\vec {p'}}
\newcommand{\vqp}{\vec {q'}}
\newcommand{\bk}{{\vec k}}
\newcommand{\bp}{{\bf p}}
\newcommand{\bq}{{\bf q}}
\newcommand{\br}{{\bf r}}
\newcommand{\bR}{{\bf R}}
\newcommand{\up}{\uparrow}
\newcommand{\down}{\downarrow}
\newcommand{\cdag}{c^{\dagger}}
\newcommand{\hlt}[1]{\textcolor{red}{#1}}
\newcommand{\ba}{\begin{align}}
\newcommand{\ea}{\end{align}}
\newcommand{\la}{\langle}
\newcommand{\ra}{\rangle}
 \title{Green's function of semi-infinite Weyl semimetals}
 
    \author{Z. Faraei}
   \affiliation{Department of Physics, Sharif University of Technology, Tehran 11155-9161, Iran}
   \author{T. Farajollahpour}
   \affiliation{Department of Physics, Sharif University of Technology, Tehran 11155-9161, Iran}

 \author{S. A. Jafari}
 \email{akbar.jafari@gmail.com}
 \affiliation{Department of Physics, Sharif University of Technology, Tehran 11155-9161, Iran}
 \affiliation{Center of excellence for Complex Systems and Condensed Matter (CSCM), Sharif University of Technology, Tehran 1458889694, Iran}
\begin{abstract}
We classify all possible boundary conditions (BCs) for a Weyl material into two classes: (i) BC that mixes the spin projection but does not change the chirality attribute, and (ii) BC that mixes the chiralities. 
All BCs are parameterized with angular variables that can be regarded as mixing angles between spins or chiralities.
Using the Green's function method, we show that these two BCs faithfully reproduce the Fermi arcs. 
The parameters are ultimately fixed by the orientation of Fermi arcs. We build on our classification and show that in the presence of a background magnetic field, only the second type BC gives rise to non-trivial 
Landau orbitals. 
\end{abstract}

\maketitle

\section {Introduction} 
Weyl equation~\cite{Weyl} describes massless relativistic particles. 
Among the elementary particles, neutrino was a candidate particle as a Weyl fermion which was nevertheless ruled out by non-zero measured mass of these particles~\cite{PRL1988}. 
Condensed matter offers a much richer platform to investigate the Weyl fermions~\cite{Bernevig,Sumathi,Burkov,Yan,Armitage,Wehling}. When the Weyl
fermions come across a boundary, they give rise to Fermi arcs -- pieces of 
constant energy surface on the boundary which do not close on themselves. 
Weyl semimetals and their associated Fermi arcs were initially proposed in pyrochlore irridates~\cite{Vishwanth2011}.
This was followed by a proposal of Burkov and Balents in heterostructures of 
topological insulators and normal insulators~\cite{BurkovBalents}. The first experimental
realization of Weyl semi-metal and Fermi arcs was on TaAs~\cite{lv2015prx,lv2015,Hassan} which
were theoretically extended within the same class of materials~\cite{TaAsClass}, including monophosphides~\cite{MonoPhos}.
Fermi arcs are observed in niobium arsenide~\cite{NbAs} as well. The possibility of
direct observation of Fermi arcs in NbP by photoemission spectroscopy remains controversial~\cite{belopolski}. 
Fermi arcs also have transport signatures. In proximity to a superconductor
there appears specific resonance in the transmission which can be regarded
as a signature of Fermi arcs~\cite{Kononov}.
In addition to condensed matter realizations,
Weyl semimetals are reported in photonic crystals~\cite{PhotonicWeyl}.

As pointed out, a key property to the Weyl semimetal phase is its unusual surface states which manifest as Fermi arcs.
An intuitive picture of Fermi arcs is as follows: They connect the two Weyl points in the Brillouin zone 
-- that always come in pairs of opposite topological charge -- by a momentum vector. Consider real space
planes specified by this momentum vector (i.e. perpendicular to this momentum). These planes will define
two-dimensional quantum spin Hall systems, the edge modes of which are topologically protected. The union
of such edge modes gives rise to a constant energy curve that connects the projection of the two Weyl nodes
on any surface in momentum space where the two projections do not coincide -- which defines the Fermi arc 
states~\cite{Ortix}. Coexistence of Fermi arcs and Dirac cone Fermi surfaces can give rise to a Lifshitz
transition in the boundary Fermi surface~\cite{Ortix}. The hallmark of Fermi arcs is their topological
protection, meaning that small disorder cannot destroy them; however, strong enough disorder dissolves
the Fermi arc into a sea of metallic bulk states~\cite{Roy}. 
The Fermi arcs can be brought together to interact in junction between Weyl 
semimetal with different Fermi arcs which then gives rise to the Fermi arc reconstruction~\cite{dwivedi}. 
In a magnetic Weyl system with open segment Fermi surface where the time reversal 
symmetry is broken, the surface plasmons become chiral Fermi arc plasmons~\cite{song}. 
Alternative picture of 
Fermi arc states can be constructed by coupling a stack of two-dimensional Fermi surfaces that
alternate between electron-like and hole-like surfaces~\cite{Hosur}.

Fermi arc states are not entirely decoupled from the bulk states~\cite{Haldane2014}.
If one separates the Fermi arc states from the bulk states, then one has to take
the coupling between them into account. Such a coupling makes the transport by Fermi 
arc states dissipative~\cite{gorbar}. This line of thinking requires to write 
a separate surface Hamiltonian capable of reproducing the Fermi arc states~\cite{Wang2017}. 
One possible method is based on the projection of the eigen-states of the bulk Hamiltonian onto 
the surface~\cite{Liu,Shan}. One can also integrate out the bulk degrees of freedom
to obtain an effective Hamiltonian for the surface states~\cite{Borchmann}.
A different effective surface Hamiltonian is used by 
Shi and coworkers to study the optical conductivity of Fermi arc states~\cite{Shi}.
In this line of thought, the mixing between the bulk and Fermi arc surface states
remains to be addressed. 

In this paper we propose an alternative approach to the Fermi arc states,
where the Fermi arcs solely arise from boundary condition (BC), and their
properties are encoded into appropriate Green's function~\cite{Tkachov,Yeyati,BursetBook}. 
In this approach one starts from the Hamiltonian for the bulk which is the Weyl Hamiltonian
composed of two chiralities $\chi=\pm 1$, and then supplements the resulting
differential equations for the Green's function with "appropriate" BCs~\cite{Arfken}.
Mathematical classification of the possible BCs gives only two type of BCs which are 
consistent with the constraint of hard wall assumption~\cite{Falko,Witten,Pinaki,Hashimoto}. 
These BCs are on the other hand consistent with the Hermiticity of the Hamiltonian~\cite{Witten}. 
The merit of the present approach is that first of all, it does not rely on any
"separation" between bulk and boundary degrees of freedom. Secondly being
based on the Green's function, it contains precise information on how do
the correlators behave as a function of the distance from the surface. This allows to theoretically
write down various quantities as an integral starting from the surface and 
reaching all the way to deep interior of the bulk. The portion of the 
contribution that scales extensively with the distance from the boundary 
will be the bulk contribution, and the rest will be exclusive due to the boundary
(Fermi arcs). After classification of the two possible BCs, we apply this 
procedure to the Landau quantization of the Fermi arc states. 


Before classification of BCs, and Landau quantization of the Fermi arcs 
using the Green's function approach, let us briefly review the literature
on the response of Fermi arcs to a background B-field. 
The Landau quantization of Weyl semimetals was experimentally measured by 
quasiparticle interference in Cd$_3$As$_2$~\cite{Yazdani}. 
The chiral anomaly of 
Weyl semi-metals is rooted in the zeroth Landau level of the bulk Weyl equation.
Direct observation of the zeroth Landau level was recently achieved by optical reflectance under high
magnetic fields~\cite{Yuan}. The semiclassical picture of the zeroth Landau level
is as follows~\cite{Potter}: The {\em chiral} zeroth Landau levels of the 
right- and left-handed Weyl fermions together with Fermi arcs on the surface
form a closed Landau orbit. Quantum oscillations associated with such a closed 
orbit will depend on the thickness of the Weyl material in the slab geometry. 
The Fermi arc surfaces alone can not host the quantum Hall effect. However, 
the "wormhole" tunneling between Weyl nodes of opposite surfaces can make it possible~\cite{Wang2017}. 

The roadmap of the paper is as follows:
in section~\ref{secII} we classify all possible boundary conditions and find
only two possibilities, one that mixes the spin directions, and the other one
that mixes the two chiralities. 
In section~\ref{GF.sec} we impose these two boundary conditions on a semi-infinite
Weyl semimetal, and establish that both BCs faithfully produce not only the Fermi arcs,
but also the dispersion relation associated with them. In section~\ref{Bfield.sec} we show
that in a background B-field, only the BC that mixes the two chiralities gives non-trivial solution,
and obtain the Fermi arc Landau levels for arbitrary Landau orbital index $n$ characterized by
decay into the interior. Our results show that the "wormhole"~\cite{Wang2017} or "conveyor belt"~\cite{Potter}
generalizes to arbitrary $n$ and is not limited to the $n=0$ level. 

\section{Classification of boundary conditions}
\label{secII}
Weyl points are points where the valence band and conduction band touch. The excitations near each Weyl point $\vec b$ are described by an effective Hamiltonian:
\bea
\hat{H}_\chi=\chi \vec \sigma \cdot (-i \vec \nabla -\vec b).
\eea
where $\chi=\pm1$ is the chirality corresponding to right-handed ($\chi=+1$) or left-handed ($\chi=-1$) fermions.
By inversion symmetry, the band touching points come in pairs, at $-\vec b$ and $\vec b$, and these have opposite chiralities. 
Note that we work in units where $\hbar= v_F=1$.  When the Fermi energy is 
at the touching points, the Weyl fermions control the low temperature physics of the solid, and novel types of surface states occur. 
So, the general equation describing Weyl fermions in the bulk is: 
\bea
\check{H}= i \hat{\tau}_z \otimes (\vec\sigma \cdot \vec \nabla) + \hat{\tau}_0 \otimes (\vec\sigma \cdot \vec b),
\eea
where the two chiralities $\chi=\pm1$ is encoded into the $\tau_z$ Pauli matrix.
This therefore naturally leads to a direct product notation which highlights the separate chirality and spin space structure.
The set of matrices $\{\hat{\tau}_x,\hat{\tau}_y,\hat{\tau}_z, \hat{\tau}_0\}$ operate in Weyl point or chirality space 
and $\{\hat{\sigma}_x,\hat{\sigma}_y,\hat{\sigma}_z, \hat{\sigma}_0\}$ act in the spin space. 
We use the notation $\check{\cal O}$ to denote the $4\times4$ matrix in the chirality$\times$spin space while
the notation $\hat{\cal O}$ applies to corresponding $2\times2$ matrix in the spin space only. 

\begin{widetext}

\begin{figure}[t]
\centering
\includegraphics[width=\linewidth]{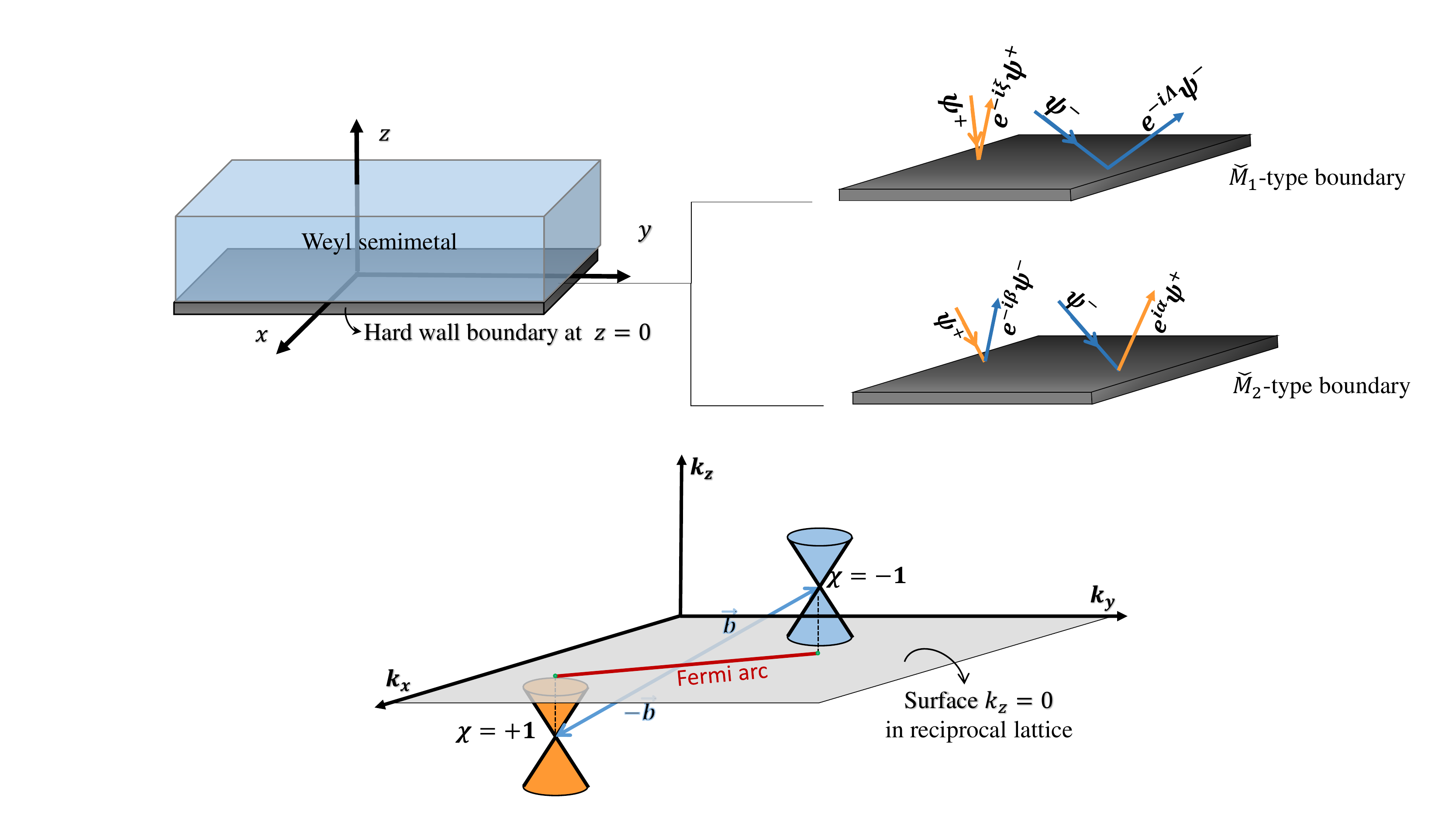}
\caption{ (Top left):Schematic representation of a semi-infinite Weyl semimetal with a hard wall boundary at $z=0$. 
(Top right): Reflection processes at the surface with $\check{M_1}$ and $\check{M_2}$ as the boundary matrix. 
The orange (blue) arrows are the electrons that come from $\chi=-1$ ($\chi=+1$) Weyl node corresponding left (right)-handed Weyl fermions. 
In both cases, the reflected electron picks up an additional phase which comes from the boundary matrix angles. These angles
turn out to be related to the slope of the Fermi arc. In the case of  $\check{M_1}$-type BC, the reflected 
electron involves spin flip to make the spin parallel to the plane of the boundary. 
In $\check{M_2}$-type BC, the reflected electron has the same spin as the incident electron, but
invloves the chirality flip such that the reflected electron is a mixture of both chiralities. 
Both types of BC are compatible with the Fermi arcs. (Bottom): Schematic illustration of the energy dispersion of the 
bulk states (Weyl cones) and the surface states, Fermi arc (red line) connecting the projections of two Weyl nodes with opposite chiralities.}
\label{Fig1}
\end{figure}
\end{widetext}

Within the geometry depicted in  Fig~\ref{Fig1}, we have a surface at position $z=0$. We are dealing with a quantum mechanical problem; namely
a Hamiltonian supplemented with "appropriate"~\cite{Witten} boundary condition. 
A "good" BC for Weyl fermions of single chirality has been discussed by Witten which immediately 
reproduces "Fermi rays" ending in the projection of the Weyl node at the boundary surface~\cite{Witten},
and is consistent with lattice models~\cite{Hashimoto}.
Incorporating the band bending near the boundary of the Weyl semi-metal turns the "Fermi rays" emanating
from the projection of the Weyl node into "Fermi spiral"~\cite{Andreev2015}. 
Accounting for simultaneous presence of two chiralities leads to an additional "good" BC. 
Details of the derivation of the most general BC which allows for possible conversion of
chiralities is discussed in appendix~\ref{app.A}. Similar considerations are applied to
cylindrical geometry~\cite{Pinaki}. 
Here is the essential line of thought of the argument:
Following Ref~\onlinecite{Falko}, a physically sensible BC is the one that prohibits 
the current transmission through the boundary. This is known as hard wall BCs that can be effectively 
incorporated into the Hamiltonian with an additional confinement potential at the boundary as,
\bea
\big[ i \hat{\tau}_z \otimes (\vec \sigma \cdot \vec \nabla) + \hat{\tau}_0 \otimes (\vec \sigma \cdot \vec b) + c \check{M} \delta(z) \big] \Psi 
= E \Psi,
\eea
where $c$ is a real constant and $\check{M}$ is a $4\times 4$, Hermitian, unitary matrix, $\check{M}^2=1$. 
Integrating the above differential equation across an infinitesimal region surrounding the boundary gives,
\bea
\big[ i \hat{\tau}_z \otimes (\vec \sigma \cdot \hat{z} )\big] \Psi |_{z=0} = c \check{M} \Psi|_{z=0}.
\eea
Substituting this equation back into itself, we find the requirements that $c^2\check{M}^2=\check{1}$.
To satisfy the above BC, it is necessary and sufficient to choose $\check{M}$ 
such that $\{ \hat{\tau}_z \otimes \hat{\sigma}_z ,\check{M} \}=0$.
There are eight possible matrices satisfying these constraints on $\check{M}$, which can be parameterized with eight parameters
as follows,
\bea
\label{bcm.eqn}
\check{M} &=& \hat{\tau}_0\otimes (a_1 \hat{\sigma}_x + a_2 \hat{\sigma}_y )+ \hat{\tau}_z\otimes (a_3 \hat{\sigma}_x + a_4 \hat{\sigma}_y)\\
\nn
&+& \hat{\tau}_x\otimes (b_1 \hat{\sigma}_0 + b_2 \hat{\sigma}_z )+ \hat{\tau}_y \otimes (b_3 \hat{\sigma}_0 + b_4 \hat{\sigma}_z).
\eea
In appendix~\ref{app.A} we show that the constraint $\check{M}\check{M}^\dag=\check{1}$ forces $\check{M}$ to have either of the following forms,
\bea
\check{M}_1 &=& \frac{\hat{\tau}_0+\hat{\tau}_z}{2}\otimes (\cos\Lambda~\hat{\sigma}_x + \sin\Lambda ~\hat{\sigma}_y )\\
\nn
&+& \frac{\hat{\tau}_0-\hat{\tau}_z}{2}\otimes (\cos\xi~ \hat{\sigma}_x + \sin\xi~ \hat{\sigma}_y)\nn\\
\check{M}_2 &=& (\cos\alpha~\hat{\tau}_x + \sin\alpha ~\hat{\tau}_y ) \otimes \frac{\hat{\sigma}_0+\hat{\sigma}_z}{2} \\
\nn
&+& (\cos\beta~\hat{\tau}_x + \sin\beta ~\hat{\tau}_y ) \otimes \frac{\hat{\sigma}_0-\hat{\sigma}_z}{2} 
\eea
where as far as the hard wall BC is concerned, the angles $\Lambda$, $\xi$, $\alpha$ and $\beta$ are some arbitrary parameters.
They describe the amount of mixing between spins ($\check{M_1}$) or chiralities ($\check{M_2}$). 
Further constraints on these parameters can be obtained from additional physical requirements. 
For example, if we consider the charge conjugation operator in the Weyl representation $C=\hat{\tau}_z \otimes \hat{\sigma}_yK$~\cite{ZeeBook} (where 
$K$ is the complex conjugation operator) and demand $C^{-1} \check{M}_1 C=\check{M}_1$, we obtain $\Lambda=\xi=\pi/2$. Similarly if we require 
$C^{-1} \check{M}_2 C=\check{M}_2$ then the boundary condition parameters will be constrained as $\beta+\alpha=\pi$.
Alternatively, they can be obtained from explicit solutions which have 
already encoded the all appropriate symmetries. This will be done in next section.

The main difference between these two BCs is that in the chirality space $\check{M}_1$ is block diagonal while $\check{M}_2$ is block-off-diagonal,
such that with $\check{M}_2$ type BC, scattering at the surface can change the chirality. Asymptotic solutions for a Weyl system with $\check{M}_2$ as the 
boundary matrix, correspond to incoming waves with one chirality that are reflected as a mixture of both chiralities. 
But $\check{M}_1$ preserves the chirality index. So, the $\check{M}$-matrix specifies the behavior of the Weyl fermions in a solid when they
hit the boundary surface. In the following, we show that either of $\check{M}_1$ or $\check{M}_2$ 
can consistently reproduce the Fermi arc. There will be one further physical piece of information
that narrows down the choice of BCs to the $\check{M_2}$. It comes from the experimental observation
of the Landau quantization of Fermi arcs~\cite{Yuan,Yazdani}. As will be shown in section~\ref{Bfield.sec}, the BC modeled by $\check{M_1}$
gives a trivial $\psi=0$ solution only. Since the two BCs can not simultaneously hold, the physical BC
can only be modeled with $\check{M_2}$ matrix. 

\section{Green's function of semi-infinite Weyl semimetal}
\label{GF.sec}
The retarded Green's function is defined by the equation,
\bea
\label{G_definition}
[\varepsilon - \check{H}(\vec r)]\check{G}(\vec r , {\vec r}~')=\delta(\vec r - {\vec r}~'), 
\eea 
where in the most general form, $\check{G}(\vec r , {\vec r}~')$ has the form:
\bea
\check{G}(\vec r, {\vec r}~')=
\begin{pmatrix}
\hat{G}_{--} & \hat{G}_{-+}\\
\hat{G}_{+-} & \hat{G}_{++} 
\end{pmatrix}.
\eea 
In the above equation $\hat{G}_{\chi\chi'}$ is of the following form,
\bea
\label{General form of G}
\\
\nn
\hat{G}_{\chi\chi'}=
\begin{pmatrix}
G_{\chi\chi'}^{\uparrow\uparrow}(z,z') & G_{\chi\chi'}^{\uparrow\downarrow}(z,z')\\
G_{\chi\chi'}^{\downarrow\uparrow}(z,z') & G_{\chi\chi'}^{\downarrow\downarrow}(z,z')\\
\end{pmatrix}
e^{[i k_x (x-x')+i k_y(y-y')]}.
\eea
The BC in terms of the Green's function $\check{G}(\vec r , {\vec r}~')$ becomes, $\check{M}\check{G}(\vec r , {\vec r}~')|_{z=0}=\check{G}(\vec r , {\vec r}~')|_{z=0}$. 
Since we consider a system that is infinite along $x$- and $y$-direction such that the momentum along the $x$- and $y$-axis are good quantum numbers, 
a plane wave part can be factorized in Eq.~\eqref{General form of G}. 

Eq.~\eqref{G_definition} implies two coupled equations for each $G_{\chi \chi'}^{\sigma \sigma'}$
with $\sigma,\sigma'$ denote the spin projections $\up,\down$, and for $\sigma=\up$, we define $\bar\sigma=\down$ and vice versa.
\bea
\label{coupled equations}
&&(\varepsilon- i \chi \sigma \partial_z +\sigma b_z)G_{\chi \chi'}^{\bar{\sigma} \sigma} - \chi (k_x^\chi + i\sigma k_y^\chi) G_{\chi \chi'}^{\sigma \sigma} =0\\
\nn
&&(\varepsilon+ i \chi \sigma \partial_z -\sigma b_z) G_{\chi \chi'}^{\sigma \sigma}- \chi (k_x^\chi - i\sigma k_y^\chi) G_{\chi \chi'}^{\bar{\sigma} \sigma} =\frac{\delta_{\chi \chi'}}{4\pi^2}\delta(z -z')
\eea
where $k_{x(y)}^\chi =k_{x(y)}-\chi b_{x(y)}$. 
Elimination of $G^{\sigma\sigma}_{\chi\chi'}$ between the coupled equations in Eq.~\eqref{coupled equations} gives,
\bea\nn
[q_\chi^2+ ( i \chi \partial_z - b_z)^2] G_{\chi \chi'}^{\bar{\sigma} \sigma }(z,z') =- \frac{k_x^\chi + i \sigma k_y^\chi}{4\pi^2 \chi} \delta(z -z')\delta_{\chi \chi'}
\eea
where $q_\chi^2=(k_x^\chi)^2 + (k_y^\chi)^2- \varepsilon^2 $. We seek the solution in the form,
\bea
\label{ansatz}
G_{\chi \chi'}^{ \bar\sigma \sigma}(z,z')&=&C_{\chi \chi'}^{ \bar\sigma \sigma}(z') e^{-(q_\chi+i\chi b_z ) z}\\
\nn
&-&\frac{ \chi(k_x^\chi + i \sigma k_y^\chi) }{8\pi^2(q_\chi+i\chi b_z)}e^{-(q_\chi+i\chi b_z) |z-z'|} \delta_{\chi \chi'},
\eea

where the first term is the solution of homogeneous part and the second one is the Green's function of the infinite system 
which is chirality diagonal (non-zero only for $\chi=\chi'$). The coefficients $C_{\chi \chi'}^{\sigma \bar{\sigma}}(z')$ are fixed by the BCs.
Once the spin-flip components of the Green's function in Eq.~\eqref{ansatz} are known, the spin-diagonal components are obtained
by Eq.~\eqref{coupled equations} which reads,
\bea
G_{\chi \chi'}^{\sigma \sigma}(z,z')&=&  \frac{\varepsilon- i \chi \sigma \partial_z + \sigma b_z}{\chi ( k_x^\chi + i\sigma k_y^\chi) } G_{\chi \chi'}^{\bar{\sigma} \sigma }(z,z').
\eea
Now let us obtain the coefficients $C^{\sigma\sigma'}_{\chi\chi'}$ for the two BCs corresponding to $\check{M}_{1,2}$
and demonstrate that in the absence of magnetic field, both BCs give rise to Fermi arcs. 

\subsection{$\check{M}_1$ as the boundary matrix}
If we choose $\check{M}_1$ as the boundary matrix, then the boundary does not mix the chiralities so we can treat them separately and solve the problem of just one chirality. 
In this situation the elements of the full Green function that mix chiralities are zero.
This leaves only $\hat G_{\chi\chi}$ to be computed which in the spin-space is,
\bea
&&\hat{G}_{\chi \chi}(\vec r , \vec r~')=
\begin{pmatrix}
1 & \frac{\varepsilon- i \chi  \partial_z +  b_z}{\chi ( k_x^\chi + i k_y^\chi) }\\
\frac{\varepsilon+ i \chi  \partial_z - b_z}{\chi ( k_x^\chi - i k_y^\chi) } & 1
\end{pmatrix}\\
\nn
&&\times
\begin{pmatrix}
0 & G_{\chi \chi}^{\uparrow \downarrow}(z,z')\\
G_{\chi \chi}^{\downarrow \uparrow}(z,z') & 0
\end{pmatrix}
e^{[i k_x (x-x')+i k_y(y-y')]}.
\eea 
The $\check{M_1}$-type BC implies,
\bea
 e^{ i \sigma\theta_\chi}G_{\chi \chi'}^{\sigma \sigma'}(z,z')|_{z=0}=G_{\chi \chi'}^{\bar{\sigma} \sigma'}(z,z')|_{z=0}
\label{bcxx.eqn}
\eea
where $\theta_-=\Lambda$ and $\theta_+=\xi$. With this BC, the coefficients of Eq.~\eqref{ansatz} are obtained as,
\bea
\label{G}
\nn
C_{\chi \chi'}^{\bar{\sigma} \sigma}&=&
\frac{\varepsilon- i \chi \sigma q_\chi + 2 \sigma b_z - \chi e^{- i \sigma \theta_\chi} (k_x^\chi + i \sigma k_y^\chi)}{\varepsilon+ i \chi\sigma q_\chi - \chi e^{- i \sigma \theta_\chi} (k_x^\chi + i\sigma k_y^\chi)}\\
&\times& \frac{\chi (k_x^\chi + i\sigma k_y^\chi)}{8\pi^2(q_\chi + i \chi b_z)}
e^{-(q_\chi+i \chi b_z)z' }\hlt{\delta_{\chi \chi'}}.
\eea

\begin{figure}[t]
\includegraphics[width=\linewidth]{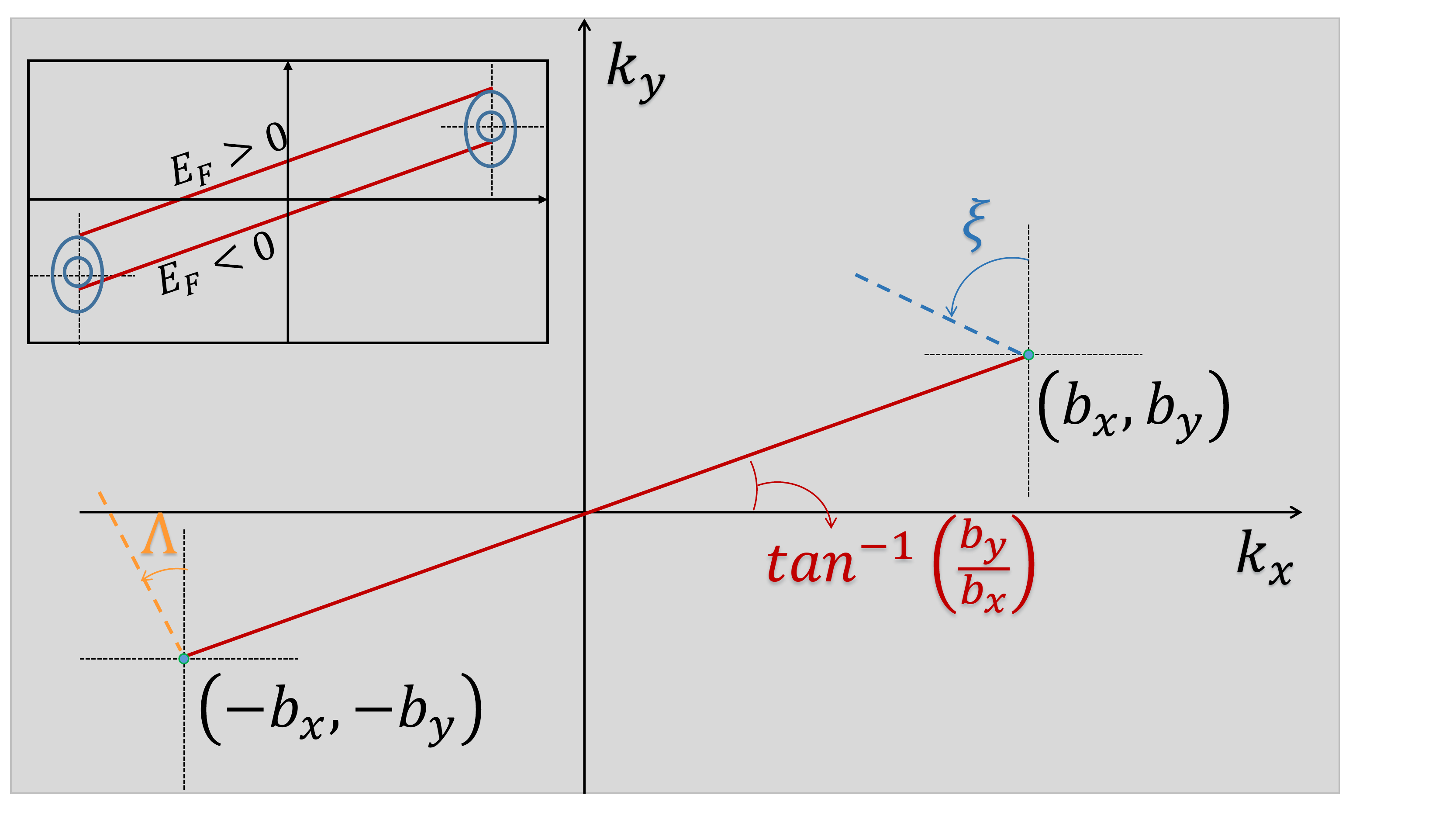}
\caption{Constant energy curves obtained from the poles of the Green's function -- Eq.~\eqref{G} -- of a confined Weyl semimetal. 
For every chirality we obtain an independent energy dispersion that terminates at the projections of the Weyl nodes on the surface of Brillouin zone. 
The slope of the Fermi rays denoted as dashed lines emenating from 
the projection of nodes are determined from the boundary matrix parameters ($\Lambda$ and $\xi$). 
In the absence of a mechanism for bending the Fermi rays, they orient themselves along the line
segment connecting the projections of the two Weyl nodes. 
Inset: When the Fermi energy is $E_F \ne 0$, the Fermi arc has a non zero intercept with $k_y$ and $k_x$ axis, consistent with the formation 
of the bulk Fermi surfaces surrounding each Weyl point~\cite{Haldane2014}. 
}
\label{Farc}
\end{figure}

The dispersion relation of Fermi arc states corresponds to the
poles of the above Green's function.
The denominator vanishes at
\bea
\label{e_M1}
\varepsilon+ i \chi\sigma q_\chi - \chi e^{- i \sigma \theta_\chi} (k_x^\chi + i\sigma k_y^\chi)=0.
\eea
This equation with the definition of $q_\chi$ as $q_\chi^2+ \varepsilon^2=(k_x^\chi)^2 + (k_y^\chi)^2 $ yields the dispersion energy of the surface states as:
\bea
\label{Fermi arc dispersion}
&&\varepsilon =\chi|k^\chi|\cos(\phi_\chi-\theta_\chi),\\
&&q_\chi= |k^\chi|\sin(\phi_\chi-\theta_\chi),
\label{q}
\eea
where $(|k^\chi|,\phi_\chi)$ are the polar coordinates corresponding to $(k^\chi_x,k^\chi_y)$ and $\theta_\chi$ as defined
above are the angles arising from BC. 
So we have found surface localized states that are supported on the {\em line segment} 
$ (k_x-\chi b_x) \cos\theta_\chi + (k_y-\chi b_y) \sin\theta_\chi- \varepsilon\chi =0$ in the $k_x-k_y$ plane. 
At $\varepsilon=0$ (which corresponds to the Fermi energy at bulk Weyl nodes), for $\chi=-1$, this is a straight line with slope of $-\cot \Lambda$ that has an end point at $(k_x,k_y)=(-b_x, -b_y)$, and for $\chi=1$, this is a line with 
the slope of $-\cot \xi$ and an end point at $(k_x,k_y)=(b_x, b_y )$. See Fig~\ref{Farc}. At the endpoints, surface localization breaks down and the solutions become plane waves with momentum $-\vec b$ for $\chi=-1$ and $\vec b$ for $\chi=1$ and so they are indistinguishable from the bulk states. 
Consequently, Eq.~\ref{Fermi arc dispersion} is a Fermi arc that has two end associated to the band crossing points $\vec b$ and $-\vec b$ and connects the projection of the Weyl points on the surface of the Brillouin zone. 

By Nielsen-Ninomiya theorem~\cite{nielsen}, the arc starting from the projection of one Weyl node has to end
in the projection of another node with the opposite chirality. Therefore the above two slopes must be
equal, i.e. $-\cot(\Lambda)=-\cot(\xi)$ which constrains the BC angles by $\Lambda-\xi=m\pi$. 
Furthermore the actual slope of the Fermi arcs is $\frac{b_y}{b_x}$ which fixes the BC angle as $\Lambda=-\cot^{-1} (\frac{b_y}{b_x})+m'\pi$.
In the above equations $m,m'$ are integers. 

For $\varepsilon \ne 0$ which corresponds to the situation where the Fermi level is shifted above or below the Weyl points, 
the line eqution acquires an intercept with respect to $(k_x,k_y)$ axis denoted in Fig.~\ref{Farc}. This result is in agreement with the 
prediction of Haldane~\cite{Haldane2014} according to which the end points of the Fermi arc are two points on the projected 
bulk Fermi surfaces (inset of Fig~\ref{Farc}). The sign of $\varepsilon$, determines the sign of the intercept. 
For $\varepsilon>0$, the shift of the end points of Fermi arcs is positive. This is satisfied if $-\frac{\varepsilon}{\sin \Lambda}>0$ 
and $\frac{\varepsilon}{\sin \xi}>0$. Therefore the sign of $\sin \Lambda$ and $\sin \xi$ are negetive and positive 
respectively, which confines $m$ to odd integers and $m'$ to even integers. 

Since we have considered only the linear corrections around the band touching points, the Fermi arc
is obtained as a straight line-segment. But in realistic materials the Fermi arcs must be bent
at least for two reasons: (i) higher order corrections taking the band curvature into account 
leads to bending in the Fermi arc which are captured in lattice models of Weyl semi-metals~\cite{Armitage}.
(ii) when the boundary is not atomically sharp, the boundary or interface can be modeled with effective
Hamiltonian which in turn gives rise to a curvature in the Fermi arc~\cite{Andreev2015,Goerbig}.
Within our picture, when a bending agent is present, the dashed Fermi rays emenating
from the projections of two nodes do not necessarily align themselves along the straight line
segment connecting the two projections. The new insight from the current analysis is that
the difference in the slope of the Fermi arc tangents at the end points of Fermi arc is
actually $\Lambda-\xi$. Therefore manipulations at the boundary are expected to alter
the curvature of the Fermi arc.

\subsection{$\check{M}_2$ as the boundary matrix}
Now if we use $\check{M}_2$-type BC, all the elements of the Green's function in the chirality space are nonzero and the BC relates the wave functions with opposite chirality:
\bea
 e^{ - i \chi\theta_\sigma}G_{\chi \chi'}^{\sigma \sigma'}(z,z')|_{z=0}=G_{\bar{\chi} \chi'}^{\sigma \sigma'}(z,z')|_{z=0},
\label{bc2.eqn}
\eea
where $\theta_\uparrow = \alpha$, $\theta_\downarrow = \beta$ and $\bar{\chi}=-\chi$. 
Imposing the $\check{M}_2$-type BC gives the coefficient $C_{--}^{\downarrow \uparrow}$ as,
\bea
&&C_{\chi \chi}^{\bar{\sigma} \sigma}
=\frac{\chi  (k_x^\chi+ i\sigma k_y^\chi)}{8\pi^2(q_\chi + i \chi b_z)} 
\big( \frac{N_{\chi \chi}^{\bar{\sigma} \sigma}}{D_{\chi \chi}^{\bar{\sigma} \sigma}} \big) 
 e^{-(q_\chi+i \chi b_z)z'}, 
\eea
where
\bea
D_{\chi \chi}^{\bar{\sigma} \sigma}&=&\bar{\chi} e^{i\chi \theta_{\bar{\sigma}}}(\varepsilon+ i \chi \sigma q_\chi ) (k_x^{\bar{\chi}} + i \sigma k_y^{\bar{\chi}})\\
\nn
&-& \chi e^{i\chi \theta_{\sigma}}(\varepsilon+ i \bar{\chi} \sigma q_{\bar{\chi}} ) (k_x^{\chi} + i \sigma k_y^{\chi}),
\eea
and
\bea
N_{\chi \chi}^{\bar{\sigma} \sigma}= D_{\chi \chi}^{\bar{\sigma} \sigma}+ 2 i  \sigma e^{i\chi \theta_{\bar{\sigma}}}  (q_\chi + i \chi b_z)   (k_x^{\bar{\chi}} + i \sigma k_y^{\bar{\chi}}).
\eea
The energy dispersion is obtained from,
\be
e^{i\chi \theta_{\bar{\sigma}}}(\varepsilon+ i \chi \sigma q_\chi ) (k_x^{\bar{\chi}} + i \sigma k_y^{\bar{\chi}})
 +
 e^{i\chi \theta_{\sigma}}(\varepsilon+ i \bar{\chi} \sigma q_{\bar{\chi}} ) (k_x^{\chi} + i \sigma k_y^{\chi})=0.\nn
\ee
This equation is the same as the previous situation Eq.~\eqref{Fermi arc dispersion}.
To see this, substitute for $\varepsilon+ i \chi \sigma q_\chi $ and $\varepsilon+ i \bar{\chi} \sigma q_{\bar{\chi}}$ from Eq.~\eqref{e_M1}, in the above
equation which then gives, $\alpha-\beta=\Lambda-\xi$ mod $2\pi$. Therefore this equation gives the
Fermi arcs in the same way that Eq.~\eqref{e_M1} does. 
Consequently, both $\check{M}_1$ and $\check{M}_2$ allow the presence of the Fermi arcs on the surface. 
Therefore as long as the experimentally observed Fermi arcs are concerned, both types of BC are compatible with the Fermi arcs.
However, when we turn on a background magnetic field to study the Landau problem of the surface states, as we will show shortly, 
the $\check{M}_1$-type BC can only produce the trivial solution $\check{\psi}=0$, while the $\check{M_2}$-type BC is compatible
with non-trivial Landau orbitals. Note that the BC proposed by Witten~\cite{Witten} only included the $\check{M_1}$-type BC which does
not give rise to Landau quantization of the surface states in Weyl materials. 

\section{Appropriate boundary conditions with background magnetic field}
\label{Bfield.sec}
A magnetic field along the $z$ direction, $\vec B= B \hat z$, can be included by minimal coupling $-i\vec \nabla\rightarrow \vec \pi=-i\vec \nabla+\frac{e}{\hbar c} \vec A$.
We work in the symmetric gauge $\vec A=\frac{1}{2}(-By,Bx,0)$. Introducing the annihilation and creation operators $a_\chi=\frac{\ell_c}{\sqrt{2}}\pi_-^\chi$ and $a_\chi^\dag=\frac{\ell_c}{\sqrt{2}}\pi_+^\chi$ to creat single particle wave functions $\phi_\chi$ of the harmonic oscillator, where $\pi_{\pm}^\chi=(\pi_x+\chi b_x) \pm i (\pi_y+\chi b_y) $ and $\ell_c=\sqrt{\frac{\hbar c}{e B}}$, $a_\chi$ and $a_\chi^\dag$ satisfy $a_\chi \phi_n^\chi=\sqrt{n} \phi_{n-1}^\chi$, $a_\chi^\dag \phi_n^\chi=\sqrt{n+1} \phi_{n+1}^\chi$ and $[a_\chi,a^\dag_\chi]=1$. In units with $\ell_c=\sqrt{2}$, the Hamiltonian can be written in terms of the $a_\chi$ and $a_\chi^\dag$ operators as,
\be
\check{H}=
\begin{pmatrix}
i\partial_z +b_z & - a_- & 0 & 0 \\
-a_-^\dag & -i\partial_z -b_z & 0 & 0 \\
0 & 0 & -i\partial_z +b_z & a_+\\
0 & 0 & a^\dag_+ & i\partial_z -b_z 
\end{pmatrix}.
\label{Hb.eqn}
\ee 
The eigen functions of this Hamiltonian are of the form 
\be
\psi_n=[f_0^n \phi_{n-1}^- ({\vec \rho}_-),f_1^n \phi_{n}^- ({\vec \rho}_-),f_2^n \phi_{n-1}^+ ({\vec \rho}_+),f_3^n \phi_{n}^+ ({\vec \rho}_+)],\label{wfn.eqn}
\ee
where $f_\mu$ with the Lorentz index $\mu=0\ldots 3$ are yet unknown functions of $z$ 
and ${\vec \rho}_\pm=\vec\rho\pm \vec b$ where $\vec \rho$ is the
transverse coordinate. Note that the center of Landau orbitals is determined by the transverse momenta, where for every Weyl point,
the momenta are measured from the projection of the bulk Weyl node on the boundary surface. 
To see the structure of the above four component spinor, first of all note that we are seeking a
separable wave function of the form $f_\mu(z) \phi_{n_\mu}(\vec \rho)$ for the Lorentz component $\mu$ where 
$n_\mu$ is its Landau level index. The above four indices in the Weyl representation break into
two sets with $\chi=\pm1$ that correspond to right-handed ($R$) and left-handed ($L$) Weyl fermions. For each chirality $\chi$,
the solutions can be written in terms of two components spinors $\varphi_1^\chi$ and $\varphi_2^\chi$ as,
\bea
\psi_n^\chi=
a_\chi
\varphi_1^\chi
e^{-iq_\chi z}
+b_\chi
\varphi_2^\chi
e^{iq_\chi z}.
\label{spinors.eqn}
\eea 
From now on we denote the four-component spinors by $\psi$, and the two components spinors (refereing
to the spin space) by $\varphi$. 
The spinors that are eigen functions of the Hamiltonian~\eqref{Hb.eqn} are given by
\be
\varphi_{1(2)}^\chi=(\phi_{n-1}^\chi ~,~\frac{\pm q_\chi+\chi \varepsilon}{\sqrt{n}}\phi_{n}^\chi)^T e^{- i\chi b_z z}.
\ee
The constants $a_\chi$ and $b_\chi$ are determined by BCs, and the chirality $\chi$ determines the center of 
the Landau orbital ($\rho_\pm$) and the relative weight of the spin $\uparrow$ and $\downarrow$. The rapidly 
varying plane wave part $e^{-i\chi b_z z}$ refers to the projection of the Weyl point with respect to which
the momenta are measured. 

Within the geometry depicted in the Fig~\ref{Fig1}, we have a surface at position $z=0$ and the Weyl material
occupies the region $z>0$. 
Let us see what could be a good BC in presence of the magnetic field $B$. 
The BC corresponding to matrix $\check{M_1}$ can be ruled out as follows:
The general form of the Landau quantized $\psi_n$ in Eq.~\eqref{wfn.eqn} 
requires the first two (as well as second two) Lorentz components to have harmonic oscillator indices $n-1$ and $n$, respectively, i.e.
they are proportional to the eigen functions $\phi_{n-1}$ and $\phi_n$ of the harmonic oscillator. 
The BC matrix $\check{M}_1$ insists to mix them. Therefore with $\check{M_1}$ type of BC, a quantized Landau level with
a definite $n$ can not be obtained. 
This leaves us with only one choice for the BC for the Landau problem of semi-infinite Weyl materials, namely $\check{M_2}$.
Fortunately the later BC matrix mixes the $\mu=0$ Lorentz component only with $\mu=2$ both of which correspond to the
harmonic oscillator index $n-1$ which are indeed the spin $\uparrow$ components of the spinor. 
Similar mixing holds for $\mu=1,3$ which correspond to harmonic oscillator index $n$ and are indeed the spin $\downarrow$
components of the spinor. This BC matrix $\check{M_2}$ is off-diagonal in the chirality space. 

Now let us see the consequences of $\check{M_2}$-type BC, $\psi_n|_{z=0}=\check{M}_2 \psi_n|_{z=0}$, which yields:
\bea
\label{boundary cond.}
&{\psi_n}_0|_{z=0}=e^{-i\alpha} {\psi_n}_{2}|_{z=0}\\
\nn
&{\psi_n}_1|_{z=0}=e^{i\beta} {\psi_n}_{3}|_{z=0}.
\eea
For incident waves of chirality $\chi=\pm$, reflection at the boundary gives the following four dimensional spinors in the chirality and spin spaces:
\bea
\label{asymp. solution}
\psi^-=\begin{pmatrix} 
\varphi_1^- \\
0
\end{pmatrix}e^{-iqz}
+ r_{-+} \begin{pmatrix}
0\\
\varphi_2^+
\end{pmatrix} e^{iqz} 
+ r_{--} \begin{pmatrix}
\varphi_2^-\\
0
\end{pmatrix} e^{iqz},\nn\\
\psi^+=\begin{pmatrix}
0\\
\varphi_1^+
\end{pmatrix} e^{-iqz} 
+ r_{+-} \begin{pmatrix}
\varphi_2^-\\
0
\end{pmatrix} e^{iqz} 
+ r_{++} \begin{pmatrix}
0\\
\varphi_2^+
\end{pmatrix} e^{iqz}.\nn\\
\eea
Here the first term of the solution $\psi^\pm$ describes an infalling wave of left($-$) or right($+$) handed electrons.
Second and third terms describe the reflected waves with opposite or the same chirality, respectively. 
Note that according to Eq.~\eqref{spinors.eqn} the infalling wave ($e^{-iq_\chi z}$) is always locked to the spinor $\varphi_1$, while
the wave reflected into the Weyl material ($e^{-iq_\chi z}$) is locked to the spinor $\varphi_2$. 
The reflection amplitudes are calculated by imposing BCs, Eq~\eqref{boundary cond.} to the most generic four-component spinor
$\psi_n$ which can be obtained by linearly combining the above left and right handed solutions. 

There are certain similarities and differences between the way the BC is imposed on three dimensional Weyl materials,
and the way the BC is imposed in two-dimensional graphene~\cite{BursetBook}. The superficial similarity between the
two cases is that in graphene we have two valley indices, while in Weyl materials we have chirality index $\chi$. 
In semi-infinite graphene with zigzag edge, the valley indices are not mixed and therefore the appropriate BC in 
our language will be given by $\check{M_1}$ matrix, while for armchair edges, the BC matrix will correspond to $\check{M_2}$. 
The essential difference between the graphene and Weyl system is that in zigzag graphene the edge contains only 
atoms of one-sublattice, say, $A$, which corresponds to only one of the pseudo-spin orientations, e.g. $\uparrow$. 
Therefore a good BC is to set the amplitude of the wave function at the other sub-lattice equal to zero which
can be separately satisfied for each of the valleys. In our case the very existence of Fermi arc states
is due to the fact that the wave function does not become zero at the boundary. Due to this difference
we need to calculate all of the components of the Green's function, $\check{G}(\vec r, {\vec r}~')=\langle \psi(\vec r)\psi({\vec r}~') \rangle$. 

Explicit form of the components of the Green's function in the spin-space will be,
\bea
\\
\nn
\hat{G}_{\chi\chi'}=
\begin{pmatrix}
G_{\chi\chi'}^{\uparrow\uparrow}(z,z')\phi_{n-1}^\chi \phi_{n-1}^{\chi'} &
G_{\chi\chi'}^{\uparrow\downarrow}(z,z')\phi_{n-1}^\chi \phi_{n}^{\chi'}\\
G_{\chi\chi'}^{\downarrow\uparrow}(z,z')\phi_{n}^\chi~ \phi_{n-1}^{\chi'} &
G_{\chi\chi'}^{\downarrow\downarrow}(z,z')\phi_{n}^\chi~ \phi_{n}^{\chi'}\\
\end{pmatrix},
\eea
where as noted earlier, the harmonic oscillator index $n-1$ ($n$) is locked to the spin index $\uparrow$ ($\downarrow$). 

For arbitrary Landau level index $n$, we can calculate the functions $G_{\chi\chi'}^{\hat{\sigma}\hat{\sigma}'}(z,z')$ from the 
differential equation that governs $\check{G}(\vec r, {\vec r}~')$. But before that, let us discuss the simplest case corresponding to the zeroth
Landau level $n=0$ which admits a much simpler solution where the coefficients in Eq.~\eqref{asymp. solution} can be separately 
obtained for each chirality $\chi=\pm 1$. 
In the lowest level we have, 
\be
\psi_0^-=\begin{pmatrix} 0\\1 \end{pmatrix} e^{i b_z z},~~~\psi_0^+=\begin{pmatrix} 0\\1 \end{pmatrix} e^{-i b_z z}, 
\ee
which represents gapless chiral modes transporting the right-handed electrons from the bulk to the surface and the reflected 
left-handed electrons from the surface to the bulk. This implies that $r_{++}=0$. Similarly $r_{--}=0$. 
So, in Eq.~\eqref{asymp. solution}, $\psi^+=(0,e^{ib_z z},0,r_{+-}e^{-ib_z z})^T$ and $\psi^-=0$. 
The BC implies that $r_{+-}=e^{-i\beta}$. This means that for $n=0$, the only possible path is that the 
right-handed electrons hitting the surface perpendicularly are reflected back into the bulk as left-handed electrons 
with phase difference $e^{-i\beta}$ with respect to the incoming electrons. 
This situation is the quantum version of the "one way conveyor-belt" 
discussed in the semi-classical treatment of Ref~\onlinecite{Vishwanth2011}. 
These authors showed that the Fermi arcs at opposite surfaces in a Weyl semimetal slab can complete the needed Fermi loop for quantum oscillations of the density of states in a magnetic field. They introduce the time of sliding of the incoming electron from '+' Weyl node along the Fermi arc before reaching the '-' Weyl node, and then they calculated the transition probability from a surface state to the bulk states. This transition probability became $\mathscr{O} (1)$, when the electron's momenta comes within $\approx \ell_B^{-1}$ of the projection of the '-' Weyl node on the surface. 
The ${\cal O}(1)$ transition probability from bulk to the Fermi arc states in the above treatment in our full
quantum mechanical treatment corresponds to the following situation: Since the only feasible BC in the
Landau problem of Weyl fermions is given by matrix $\check{M_2}$, the wave function (four component spinor) of the $n=0$ Landau level 
at the boundary is forced by BC to exchange the chiralities with exact probability of $1$. 
Therefore our quantum mechanical treatment, sharpens the ${\cal O}(1)$ transition probability of right and
left chiralities into exactly $\equiv 1$ probability for the $n=0$. 
The above semi-classical treatment is valid for weak enough magnetic
fields where the Fermi arc itself still exists, while our argument holds for arbitrary $B$. 
In the next section we consider higher Landau levels with $n>0$.

\section{Green's function in the presence of a mgnetic field}
\label{with magnetic field}
The retarded Green's function satisfies the equation
\bea
[\varepsilon - \check{H}(\vec r)]\check{G}(\vec r , {\vec r}')=\delta(\vec r - {\vec r}~').
\eea 
In terms of $\check{G}(\vec r , {\vec r}')$, the boundary condition reads $\check{M}_2\check{G}(\vec r , {\vec r}')|_{z=0}=\check{G}(\vec r , {\vec r}')|_{z=0}$. Eq.\eqref{G_definition} implies two coupled equations for each $G_{\chi \chi'}^{{\sigma} {\sigma}'}$: 
\bea
&&-\chi \sqrt{n} G_{\chi \chi'}^{\sigma \sigma} + (\varepsilon- i \chi \sigma \partial_z +\sigma b_z)G_{\chi \chi'}^{\bar{\sigma} \sigma}=0\\
\nn
&&(\varepsilon+ i \chi \sigma \partial_z -\sigma b_z) G_{\chi \chi'}^{\sigma \sigma}- \chi \sqrt{n} G_{\chi \chi'}^{\bar{\sigma} \sigma} =\delta(\vec r -{\vec r}~')\delta_{\chi \chi'}
\eea
which yields:
\bea\nn
[q^2+ ( i \chi \partial_z - b_z)^2] G_{\chi \chi'}^{\bar{{\sigma}} {\sigma} }(z,z') =- \chi \sqrt{n} \delta(z -z')\delta_{\chi \chi'},
\eea
where $q^2=n - \varepsilon^2 $ aand $\bar{{\sigma}}=-{\sigma}$. We seek the solution in the form,
\bea
\label{cudb.eqn}
G_{\chi \chi'}^{ \bar{{\sigma}} {\sigma}}(z,z')&=&C_{\chi \chi'}^{ \bar{{\sigma}} {\sigma}}(z') e^{-(q+i\chi b_z ) z}\\
\nn
&-&\frac{ \chi \sqrt{n} }{2(q+i\chi b_z)}e^{-(q+i\chi b_z) |z-z'|} \delta_{\chi \chi'}, \\
\nn\\
\nn
G_{\chi \chi'}^{{\sigma} {\sigma}}(z,z')&=& \chi \frac{\varepsilon+ \bar{{\sigma}} ( i\chi \partial z - b_z)}{\sqrt{n}} G_{\chi \chi'}^{\bar{{\sigma}} {\sigma} }(z,z'),
\eea
where the first term is the solution of homogeneous part and the second one is the Green's function of the 
infinite system for $\chi=\chi'$. The coefficients $C_{\chi \chi'}^{{\sigma} \bar{{\sigma}}}(z')$ are obtained from
Eq.~\eqref{bcxx.eqn} as,
\bea
\nn
C_{\chi \chi}^{\bar{\sigma} \sigma}&=&
\frac{(\varepsilon - i \chi \sigma q)+ e^{- i \chi \sigma(\alpha + \beta)}(\varepsilon - i \chi \sigma q + 2 \sigma b_z)}{(\varepsilon- i \chi \sigma q)+ e^{- i \chi \sigma(\alpha + \beta)}(\varepsilon + i \chi \sigma q )}\\
&\times& \frac{\chi \sqrt{n}}{2(q+i \chi b_z)}
e^{-(q+i \chi b_z)z'}
\eea
The poles are given by, $(\varepsilon- i \chi \sigma q)+ e^{- i \chi \sigma(\alpha + \beta)}(\varepsilon + i \chi \sigma q )=0$. So the dispersion energy of the surface states in the presence of a magnetic field perpendicular to the surface is $\varepsilon=\pm \sqrt{n} \sin (\frac{\alpha+\beta}{2})$.
As $\varepsilon\rightarrow \pm \sqrt{n} \sin (\frac{\alpha+\beta}{2})$, surface Green's function behaves as:
\bea
&&G_{\chi \chi'}^{{\sigma}' {\sigma}} (z,z')=
\frac{-{\sigma} \sqrt{n}}{\varepsilon+\sqrt{n}\sin(\frac{\alpha+\beta}{2})} 
\bigg( \frac{e^{-q(z+z')}}{1+e^{-\chi{\sigma}(\frac{\alpha+\beta}{2})}} \bigg)\\
\nn\\
\nn
&&\times
\left \{
\begin{array}{ll}
\chi e^{- \sigma \chi i (\frac{\alpha+\beta}{2})} e^{-i \chi b_z (z+z')} & \text{if}~\sigma '=\sigma\\
ie^{-\sigma\chi i (\alpha+\beta)} e^{-i\chi b_z (z+z')} & \text{if}~ \sigma '=-\sigma~\text{and}~\chi'=\chi\\
ie^{\chi i \theta_\sigma} e^{-i\chi b_z (z-z')} & \text{if}~ \sigma '=-\sigma~\text{and}~\chi'=-\chi
\end{array}
\right. .
\eea
This is the spectrum of the surface states, decaying exponentially from the surface $z=0$. 

\section{Summary}
We derived all possible hard wall BCs for Weyl semimetals. We classified the BCs into two 
generic types that mix either the spins or the chiralities. Mathematically, each of the BCs is parameterized
by two angular variables. This is enough to faithfully reproduce the Fermi rays emanating from
the projection of a given Weyl node on the surface of the BZ. The slope of each Fermi ray is
characterized by either of the two parameters. Requiring the rays to merge into a line-segment
(Fermi arc), completely fixes the angular variables. Adding a background B-field, only 
second type of BC that allows the chiralities to mix (leaving the spins intact) will 
give rise to non-trivial solutions. 

Building on the above BCs, we employed the powerful Green's function method
to first of all reproduce the well known results on the Fermi arcs of the Weyl
semi-metals. Then we studied the Landau quantization with this technique. 

The advantage of the present Green's function approach is that it involves no
separation of bulk and surface degrees of freedom, and all the degrees of 
freedom are taken into account on the same footing. While in approaches that
separate the bulk and boundary degrees of freedom, at the end one needs to
address the mixing between the two, in the present approach, a continuous cross-over
from bulk to boundary degrees of freedom is automatically built into the Green's function.
The part of the Green's function that exponentially decays towards the interior is due
to the boundary (Fermi arc) degrees of freedom. Those parts that do not decay by
moving away from the boundary contribute extensively to physical properties. 

The chiral anomaly which is a hallmark of Weyl semi-metals rests on a 
mechanism to convert the two chiralities to each other. Then a background
electromagnetic field in principle can take advantage of this possibility
to generate chirality imbalance which in turn gives rise to chiral anomaly. 
Our consideration of the BCs, suggests that the $\check{M_2}$-type BC allows
conversation between the chiralities. This suggests that an arrangements of the
electromagnetic fields in the boundary can also induce a chiral anomaly.



\appendix
\section{M Matrix \label{app.A}}
In this appendix we parameterize the most general form of matrices, $\check{M_1}$ and $\check{M_2}$
which are allowed by requiring that there is no electron current coming out of the Weyl material. 
The matrix $\check{M}$ in Eq.~\eqref{bcm.eqn} can be explicitly written as,
\begin{widetext}
\bea
\check{M}=
\begin{pmatrix}
0 & (a_1+a_3) - i(a_2+a_4) & (b_1+b_2)-i(b_3 +b_4) & 0\\
(a_1+a_3) + i(a_2+a_4) & 0 & 0 & (b_1-b_2)-i(b_3 - b_4) \\
(b_1+b_2)+i(b_3 +b_4) & 0 & 0 & (a_1-a_3)-i (a_2-a_4)\\
0 & (b_1-b_2)+ i(b_3 - b_4) & (a_1-a_3)+i (a_2-a_4) & 0
\end{pmatrix}.
\eea 
\end{widetext}
Reparameterizing in an obvious manner we have,
\bea
\begin{pmatrix}
0 & x-iy & t-iq & 0\\
x+iy & 0 & 0 & r-is \\
t+iq & 0 & 0 & u-iv\\
0 & r+is & u+iv & 0
\end{pmatrix}.
\eea 
The diagonal elements of $\check{M}^2=1$ give,
\bea
(x^2+y^2)+(t^2+q^2)=1,\\
\nn
(x^2+y^2)+(r^2+s^2)=1,\\
\nn
(u^2+v^2)+(r^2+s^2)=1,\\
\nn
(u^2+v^2)+(t^2+q^2)=1,
\eea
which allows for angular reparameterization as,
\bea
x=\cos \Lambda \cos \gamma~,~ y=\sin \Lambda \cos \gamma,\\
\nn
u=\cos \xi \cos \gamma~,~ v=\sin \xi \cos \gamma,\\
\nn
t=\cos \alpha \sin \gamma~,~ q=\sin \alpha \sin \gamma,\\
\nn
r=\cos \beta \sin \gamma~,~ s=\sin \beta \sin \gamma.
\eea
The off-diagonal components of $\check{M}^2=1$ give the further constraints,
\bea
yr+xs+qu+tv=0,\\
\nn
xt+qy+ru+sv=0,\\
\nn
xr-ys+tu-qv=0,\\
\nn
yt-xq+rv-su=0.
\eea
There are three ways to satisfy them simultaneously:
\begin{itemize}
\item 
If $\gamma=0$ so $t=q=r=s=0$, $x \pm i y=e^{\pm i \Lambda}$ and $u \pm i v = e^{\pm i \xi}$ and 
\bea
\check{M}=\check{M}_1=
\begin{pmatrix}
0 & e^{- i \Lambda} & 0 & 0\\
e^{ i \Lambda} & 0 & 0 & 0 \\
0 & 0 & 0 & e^{- i \xi}\\
0 & 0 & e^{ i \xi} & 0
\end{pmatrix}.
\eea
\item If $\gamma=\pi/2$ so $x=y=u=v=0$, $ t\pm iq=e^{\pm i \alpha}$ and $r \pm i s = e^{\pm i \beta}$ and 
\bea
\check{M}=\check{M}_2=
\begin{pmatrix}
0 & 0 & e^{- i \alpha} & 0\\
0 & 0 & 0 & e^{- i \beta} \\
e^{ i \alpha} & 0& 0 & 0 \\
0 & e^{ i \beta} & 0 & 0
\end{pmatrix}.
\eea
\item If $(\Lambda-\xi)+(\beta-\alpha)= m\pi$, with $m$ an integer we obtain, 
\bea
\check{M}=\check{M}_3= (\cos \gamma) \check{M}_1 +(\sin \gamma) \check{M}_2.
\eea
\end{itemize}
The last possibility is ruled out by $M=M^\dagger$. We are therefore left with only 
$\check{M_1}$ and $\check{M_2}$. The $\check{M_1}$ is chirality-diagonal. It only 
linearly combines the transverse spin components $\sigma^x,\sigma^y$ with angles
$\Lambda$ and $\xi$ for the left and right chiralities, respectively. The second
choice, $\check{M_2}$ is diagonal in the spin space, meaning that it does not
alter the spin of the incident electron, while it allows to flip the chirality
as it is off-diagonal in the chirality space.

\bibliographystyle{apsrev4-1}
\bibliography{Refs}

\end{document}